# Link and Network-wide Study of Incoherent GN/EGN Models

Farhad Arpanaei[(1)], M.Ranjbar Zefreh[(2)], Jose A.Hernandez[(1)], Andrea Carena[(3)], David Larrabeiti[(1)]

[(1)] Universidad Carlos III de Madrid, Departamento de Ingeniería Telemática, farhad.arpanaei@uc3m.es
[(2)] Cisco Systems Italy S.r.l,
[(3)] Politecnico di Torino, Dipartimento di Elettronica e Telecomunicazioni.

**Abstract** *An unprecedented comparison of closed-form incoherent GN (InGN) models is presented with heterogeneous spans and partially loaded links in elastic optical networks. Results reveal that with accumulated dispersion correction and modulation format terms, the InGN shows higher accuracy.*
©2022 The Author(s)

**Introduction**
Among several proposed approaches for the non-linearity assessment of coherent multi-span optical fiber systems, four approaches have been more popular in estimating signal-to-noise ratio (SNR) [1-7]. According to descending accuracy and run-time, these models are 1- Split Step Fourier Method (SSFM) [4], 2- Integral-based Enhanced Gaussian Noise (EGN) model [5], 3- Integral-based Gaussian Noise model [6], and 4- Closed-form Models (CFMs) [7]. While SSFM is a time-domain method, the other approaches mentioned above are based on frequency domain analysis [4]. The accuracy of SSFM and integral-based EGN is similar and very close to the experimental test results. However, in integral-based GN, due to neglecting the effect of modulation format levels (MFLs) of the channel under test (CUT) and the interfering channels, we have some errors concerning accurate EGN results, although the amount of the error decreases as the length of the link increases [6]. Despite the excellent accuracy of SSFM and Integral-based GN/EGN models, since the non-linear interferences (NLIs) of different spans are coherently calculated in GN/EGN integral form methods they are not applicable for network planning with add/drop ability at the intermediate nodes due to the time necessary for numerical calculation of the integrals. Concerning these issues, some closed-form models (CFMs) have been presented by analytic approximation of the integrals in GN and EGN methods [6]. The authors in [7] proposed several versions of CFM for the C-band, including versions from CFM0 up to CFM4. Indeed, CFM0-2 are incoherent models that can be used in network-wide studies. It should be noted that CFM2 includes the MFL and accumulated dispersion correction terms of both the CUT as well as other interfering channels. Thus, from the five versions presented in reference [7], we only focus on CFM2, the most accurate version among incoherent versions aimed in this paper. Another alternative is CFM presented in [8] (the correction version is presented in [9]), where the authors introduced a CFM with an MFL correction term for interfering channels only. The CFM in [9] is applicable both for C and C+L bands, and we consider the C-band version (simply by turning off the Raman effect in equation 3 in [9]) for the sparsely loaded links and heterogenous spans. Finally, we modify the CFM for heterogeneous spans with the MFL correction term of interfering channels introduced in equation 7.32 from chapter 7 in [10]. Therefore, to the best of the authors' knowledge, this is the first study to compare CFM models with and without MFLs and accumulative dispersion terms at the link and network-level for elastic optical networks.

**System Model and Methodology**
To have a comprehensive picture, we evaluate five CFMs, including 1) without MFL and accumulated dispersion correction terms CFM (CFM0 in [7]) called WoMDCT-1 in this paper, 2) modified version of CFM without MFL and accumulated dispersion correction terms CFM for C-band in [9] called WoMDCT-2 in this paper, 3) modified version of CFM including MFL but without accumulated dispersion correction terms CFM in [10] called MCT-1 in this paper, 4) modified version of CFM with MFL but without accumulated dispersion correction terms CFM in [9] called MCT-2 in this paper, and finally 5) CFM with MFL and with accumulated dispersion correction terms (CFM2 in [7]) which is called MDCT in this paper. We first perform a link-level study to evaluate the accuracy of different CFMs. Then, we will present a network-wide study to compare the bandwidth blocking probability, GSNR, and MFL usage for the winner CFM in each class. 500 partially loaded PtP links are generated using 100 heterogeneous spans, i.e., spans with different random lengths in a link-level analysis. Based on the Local Optimization Global Optimization for Nyquist waveforms called LOGON in [1], the optimum launch power is estimated for each sample. Moreover, the number of spans has been refined according to the MFL, randomly selected for each sample from 1 to 6 according to the PM-BPSK, PM-QPSK, PM- 8, 16, 32, and 64QAM. From Eq.1, we calculate the generalized SNR (GSNR) of a lightpath (LP) 's CUT by applying the CFMs [2,11].

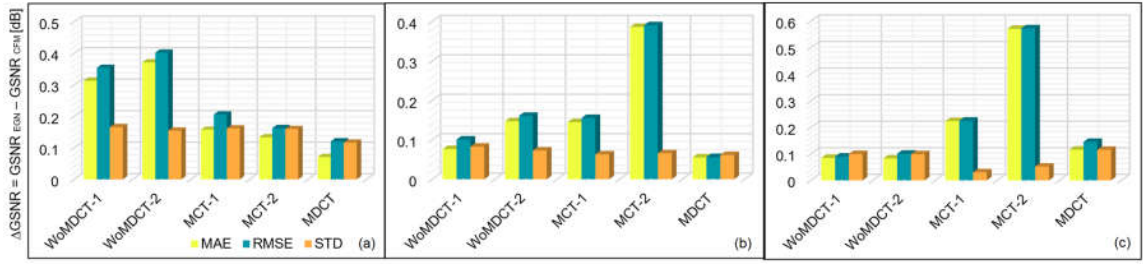

**Fig. 1:** $\Delta GSNR = GSNR_{EGN} - GSNR_{CFM}$ for point-to-point partially loaded links with heterogonous spans (a) 500 links with modulation format levels (MFLs): PM- 8, 16, 32, and 64QAM, (b) 5 links with MFL: PM-QPSK, and (c) 5 LPs with MFL: PM-BPSK.

$$GSNR_{LP}^{m,r} \approx \frac{P_{ch}^{m,l,s} - \sum_{l=1}^{L^r} \sum_{s=1}^{S^{r,l}} P_{NLI}^{m,l,s}}{\sum_{l=1}^{L^r} \sum_{s=1}^{S^r} P_{ASE}^{m,l,s} + \sum_{l=1}^{L^r} \sum_{s=1}^{S^r} P_{NLI}^{m,l,s}} \quad (1)$$

where $m, r, l,$ and $s$ are the indices of $m^{th}$ channel, $r^{th}$ LP, $l^{th}$ link, and $s^{th}$ span, respectively. Moreover, $P_{NLI}^{m,l,s}$ and $P_{ASE}^{m,l,s}$ are the noise power of NLIs and erbium-doped fiber amplifiers of each span. $P_{NLI}^{m,l,s}$ can be obtained from modified versions of equations as mentioned above in [7], [9], and [10] for five CFMs related to the sparsely loaded links and heterogeneous spans. Moreover, $P_{ASE}^{m,l,s} = hf^m N_F^{l,s}(G_{span}^{l,s} - 1)$, where $G_{span}^{l,s}$ is the amplifier gain of span $s$ on link $l$. Moreover, we assume link transparency, i.e., $G_{span}^{l,s} = e^{2\alpha^{l,s} L_{span}^{l,s}}$, that is, that the span amplifier exactly compensates for the loss of each span. Indeed, the span length may be different. In addition, $L_{span}^{l,s}$ is the length. Our baseline to compare all CFMs is the results of the EGN emulator created based on the Integral-based EGN model's equations [5]. Since the run time of long-haul PtP links (with over 40 spans), such as those with PM-BPSK and PM-QPSK, is usually time-consuming [7], we evaluated only five samples for each MFL in those cases. Furthermore, we generated 500 samples of PtP links with PM- 8, 16, 32, and 64QAM.

**Simulations and Results**
In the link-level study, each sample includes 12 attributes, i.e., the channels' launch power, the number of channels in the link ($N_{ch}^{l,s}$), the fiber field loss, the dispersion, and the fiber non-linearity coefficients of span $s$, channel spacing, symbol rate, and frequency center of channel $m(n)$ ($f^{m(n)}$), MFL correction factor of a channel [7], amplifier's noise figure, and the length and number of spans. Eq.(1) is also used for link-level study, where the LP has a PtP link. In this study, to generate 510 (500+5+5) PtP links samples, we assume the busy channels' symbol rates and their bandwidths are equal to 64 GBaud and 75 GHz (6x12.5 GHz), respectively. Thus, $N_{ch}^{l,s}$ = 60 in C-band, i.e., $f^{m(n)} \in \{191.61, \ldots, 195.95\}$ THz. Also, the fiber field loss, the dispersion, and the fiber non-linearity coefficients of span $s$, and amplifier's noise figure are 0.21 dB/km, -21.45x$10^{-27}$ s$^2$/m, 1.31x$10^{-3}$ (W.m)$^{-1}$, and 6 dB, respectively. Furthermore, channels' launch power are selected from [-5,5] dBm with resolution 0.01 dBm. Moreover, the length and number of spans are selected from [50,120] km, and equals 100, respectively. Also, BER$_{threshold}$ = 3.8x$10^{-3}$ is suitable for a 28% forward error correction overhead. Thus, according to equations in [5], the GSNR thresholds for MFLs: 1-6 (i.e., PM- BPSK, PM-QPSK, PM-8QAM, PM-16QAM, PM-32QAM, and PM-64QAM, respectively) are 5.52, 8.53, 12.51, 15.19, 18.19, and 21.12 dB, respectively. Note the number of spans is initially considered large enough. Afterward, we refine it by calculating the GSNR according to the BER and related MFLs' required GSNRs. In this paper, we did not consider the aging margin to have a fair comparison of CFMs. Moreover, the CUT for each sample was randomly selected among the busy channels. The loading status of each sample is randomly selected from [10,100] (in percent), showing what percentage of the link's channels are busy. In Fig.1, the GSNR deviations ($\Delta GSNR = GSNR_{EGN} - GSNR_{CFM}$) are illustrated for five CFMs in three classifications in MFLs. For MFL: 3-6 (Fig.1(a)), the MDCT and MCT-2 present accurate results. However, the merit root mean square error (RMSE) and mean absolute error (MAE) of MDCT is lower than MCT-2. The RMSE and MAE of MDCT are 0.05 and 0.1 dB, respectively. The standard deviation (STD) of results for MDCT is about 0.1 dB, which shows the consistent accuracy of the MDCT. Interestingly, as depicted in Fig.1(b) and Fig.1 (c), the measurement merits of WoMDCTs and MDCT are better than MCTs for PM-BPSK and

PM-QPSK. The main reason is that the GSNRs of long-haul PtP links obtained based on WoMDCTs and MDCT with approximately non-Gaussian signals converge when the links have large number of spans. Thus, to have consistent accuracy for long and short PtP links, we should simultaneously consider the MFL and dispersion correction terms of CUT and interfering channels. In the following network-wide study, we compare three winner CFMs according to our classification, i.e., WoMDCT-1, MCT-2, and MDCT. Having showed that proposed models are more accurate, we carried out a blocking probability study on two network topologies to quantify the impact of the improved physical layer modelling. We consider two topologies: a regional network, i.e., Italy's national backbone (ITB) (21 nodes, 36 links, average link length of 166 km) [12], and a long-haul network, i.e., the US national backbone (USB) (24 nodes, 43 links, average link length of 1000 km) [13]. The bit-rates supported by transceivers are a function of the applied MFLs and are equal 92, 184, 276, 368, 460, and 552 Gb/sec, respectively. In Tab.1, the maximum reach distance and transmission reaches of MFLs are reported for selected CFMs. The results show that using each CFMs can affect the span count and optimum powers. If the requested bit rate exceeds the maximum capacity of a single transceiver using a given MFL, the request is established on several carriers using the same MFL, transmitted within one spectral SbCh, allocated on the adequate number of adjacent slices. Traffic requests are generated randomly, with the bit-rates between 200 Gb/s and 1 Tb/s, with a 200 Gb/s resolution. The offered traffic load (OTL) is AT/HT normalized traffic units [13], where the requests arrive according to a Poisson process with an average arrival rate of AT requests per time unit. Also, the holding time of each request is generated according to a negative exponential distribution with an average of 1/HT. Each OTL value is simulated with $10^5$ requests repeated five times in each OTL. We use $k = 3$ shortest paths applied to generate candidate LPs. As shown in Fig.2, the bandwidth blocking probability (BBP) for MDCT in both networks is lower than other CFMs, since LPs with the same distance MCDT use transceivers with higher MFLs. For example, the usages of PM-64QAM in ITB averagely are 34%, 23%, and 14% over OTLs for MDCT, MCT-2, and WoMDCT-1, respectively, while PM-BPSK and PM-QPSK are never used. The same behaviour regarding the PM-64QAM is seen in the USB with WoMDCT-1. Additionally, since the USB averagely established LPs larger than ITB, the BBP in the same OTL is higher for the USB. Furthermore, we find that applying MDCT CFM about 1 dB improves the mean GSNR of established LPs at the arrival time in two networks (see Fig.3).

**Tab. 1:** The maximum reach distance in spans counts for span length equals 80 km.

| MFL | 1 | 2 | 3 | 4 | 5 | 6 | Launch Power [dBm] |
|---|---|---|---|---|---|---|---|
| | Maximum Reach Distance in terms of span number | | | | | | |
| WoMDCT-1 | 140 | 70 | 28 | 15 | 7 | 3 | 1.44 |
| MCT-2 | 149 | 74 | 30 | 16 | 8 | 4 | 1.69 |
| MDCT | 189 | 94 | 37 | 20 | 10 | 5 | 1.09 |

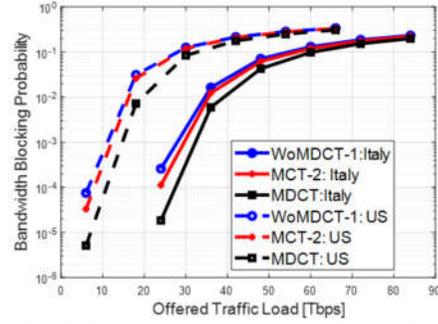

**Fig. 2:** Bandwidth blocking probability v.s. offered traffic load for different GSNR closed-forms.

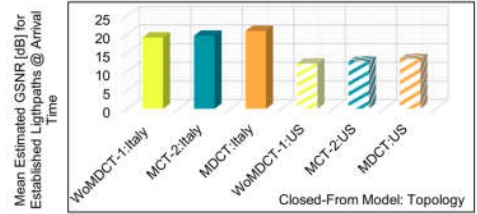

**Fig. 3:** Mean arrival time GSNR of established LPs for ITB and USB with different GSNR closed-forms.

## Conclusions
The results reveal that the CFM with modulation format level and dispersion correction terms can reduce the SNR design margin and channel launch power by approximately 1 dB and 0.5 dBm, respectively. We observe a decrease of blocking probability because the previously considered CFMs were overestimating non-linear interference.

## Acknowledgements
Farhad Arpanaei acknowledges support from the CONEX-Plus programme funded by Universidad Carlos III de Madrid and the European Union's Horizon 2020 research and innovation programme under the Marie Sklodowska-Curie grant agreement No. 801538. The authors would like to acknowledge the support of the EU-funded B5G-OPEN project (grant no. 101016663) and the Spanish projects ACHILLES (PID2019-104207RB-I00) and Go2Edge (RED2018-102585-T). This work was supported by the Italian Ministry for University and Research (PRIN 2017, project FIRST).